\documentclass[namedreferences]{solarphysics}
\usepackage[natbib,optionalrh,solaenum]{spr-sola-addons} 
\usepackage{graphicx}                    
\usepackage{color}                       
\usepackage[urlcolor=blue,breaklinks=true]{hyperref}
\usepackage{breakurl}                         

\usepackage{threeparttable}	
\usepackage{adjustbox}

\newcommand{\lya}{Ly$\alpha$}
\newcommand{\lyb}{Ly$\beta$}
\newcommand{\lyc}{Ly$\gamma$}

\begin{document}
\begin{article}

\begin{opening}

\title{Extreme Ultra-Violet Spectroscopy of the Lower Solar Atmosphere During Solar Flares}

\author{Ryan~O.~\surname{Milligan}$^{1,2,3}$\sep}

\runningauthor{Milligan}
\runningtitle{EUV Spectroscopy of the the Lower Solar Atmosphere During Solar Flares}

\institute{$^{1}$ Astrophysics Research Centre, School of Mathematics and Physics, Queen's University Belfast, University Road, Belfast, BT7 1NN, Northern Ireland \\
              $^{2}$ Solar Physics Laboratory, Heliophysics Division, NASA Goddard Space Flight Center, Greenbelt, MD 20771, USA \\
              $^{3}$ Department of Physics Catholic University of America, 620 Michigan Avenue, Northeast, Washington, DC 20064, USA \\
              email: \url{r.milligan@qub.ac.uk} \\ 
              }

\begin{abstract}
The extreme ultraviolet (EUV) portion of the solar spectrum contains a wealth of diagnostic tools for probing the lower solar atmosphere in response to an injection of energy, particularly during the impulsive phase of solar flares. These include temperature and density sensitive line ratios, Doppler shifted emission lines and nonthermal broadening, abundance measurements, differential emission measure profiles, and continuum temperatures and energetics, among others. In this paper I shall review some of the recent advances that have been made using these techniques to infer physical properties of heated plasma at footpoint and ribbon locations during the initial stages of solar flares. I shall primarily focus on studies that have utilized spectroscopic EUV data from Hinode/EIS and SDO/EVE as well as providing some historical background and a summary of future spectroscopic instrumentation.
\end{abstract}


\end{opening}

\section{Introduction} 
\label{s:Introduction} 
The spectroscopy of extreme ultra-violet (EUV) emission is a crucial diagnostic tool for determining the composition and dynamics of the flaring solar atmosphere in response to an injection of energy. It is well known that the chromosphere\footnote{I shall use the terms chromosphere, lower transition region, and lower solar atmosphere interchangeably, since we do not know how to adequately describe this region during a flare's impulsive phase. In essence I am referring to the footpoints or ribbons of a flaring loop, which are often spatially coincident with HXR emission. This may also include the photosphere in the case of white light flares.} is the dominant source of radiated energy during a solar flare \citep{neid89,huds92,huds06,kret10,kret11,wood04,wood06}, as well as being the origin of material that is ultimately observed in the corona through the process of chromospheric evaporation \citep{neup68}. But how the energy initially stored in the coronal magnetic field gets transported to the chromosphere, where it drives increased emission across a range of wavelengths, remains poorly understood. It is commonly assumed that the delivery mechanism is a `beam' of accelerated (nonthermal) electrons since enhanced chromospheric (and occasionally photospheric) emission is temporally and/or spatially correlated with hard X-ray (HXR) emission due to thick-target bremsstrahlung. However other processes such as thermal conduction fronts, ion beams, Alfv\'{e}n waves, return currents, etc., may also play a role. While imaging instruments provide important context information on the morphology and structure of solar features, the images themselves are usually broadband, comprising several different ion species that can bias the interpretation of the observations. Spectroscopy offers the advantage of providing quantifiable measurements of parameters such as temperature, density, and velocity that can then be compared with predictions from theoretical models.

The EUV portion of the solar spectrum spans the 100--1200\AA\ wavelength range \citep{tobi06}, although some in the aeronomy discipline use 300\AA\ as the lower limit due to instrumental factors \citep{wood05}, while others also include the prominent Lyman-alpha (\lya) line at 1216\AA\ \citep{lile08}, the strongest line in the solar spectrum (see Section~\ref{lya_flares} for a discussion on \lya\ flare observations). The most prominent chromospheric features in the EUV range are the free-bound continua of H~I and He~I, and to a lesser extent, He~II, with recombination edges at 912\AA, 504\AA, and 228\AA, respectively, as well as the resonance line of He~II at 304\AA. During large flares free-free (thermal bremsstrahlung) continuum emission from the corona can also encroach into the EUV spectral range \citep{mill12a,mill13}. However, the EUV range also contains myriad emission lines with intrinsic formation temperatures from 10$^{4}$~K, which are characteristically chromospheric, up to a few times 10$^{7}$~K that are only found during flares or in the cores of large active regions. It is when this inherently coronal emission of perhaps a few megaKelvin is observed to be spatially and/or temporally correlated with enhanced chromospheric emission that we are witnessing the direct manifestation of flare heating. Analysis of these lines and continua can therefore allow us to deduce the state of the heated plasma. Energetically speaking, even the strongest chromospheric EUV features contain, at most, a few per cent of a flare's total energy \citep{emsl78b,mill12a,mill14}, however the diagnostic information contained therein is invaluable for quantifying changes in the plasma parameters in response to injected energy.

The EUV component of radiated emission is also known to be a major energy input into the Earth's upper atmosphere and the geospace environment, heating the thermosphere and generating the ionosphere \citep{lean09,qian10}. In particular, He~II 304\AA\ photons are the most important energy contributor to overall heating in the thermosphere at all levels of solar activity, while chromospheric \lya\ emission (along with soft X-rays; SXR) generates the D-layer of the ionosphere (80--100~km; \citealt{tobi00}). The bulk of the Sun's EUV emission, however, gets absorbed by the F-layer ($>$150~km), while the E-layer (100--150~km) is affected by X-ray Ultra-Violet radiation (XUV; 1--100\AA). Changes in the Sun's output at EUV wavelengths, particularly during flares, can drive fluctuations in the ionospheric density, which increases drag on orbiting satellites, disrupts long-wave radio communication and compromises GPS accuracy. The EUV component of radiation emitted during the early stages of solar flares often originates in the chromosphere. The study of chromospheric EUV emission is therefore not only important for understanding flare heating, but also in determining the origins of a major component of space weather.

Despite the amazing diagnostic potential currently available to us through instruments such as the EUV Imaging Spectrometer (EIS; \citealt{culh07}) onboard Hinode \citep{kosu07} and the EUV Variability Experiment (EVE; \citealt{wood05}) on the Solar Dynamics Observatory (SDO; \citealt{pesn12}) - and previously from the Coronal Diagnostic Spectrometer (CDS; \citealt{harr95}) and Solar Ultraviolet Measurements of Emitted Radiation (SUMER; \citealt{wilh95}) instruments onboard the Solar and Heliospheric Observatory (SOHO) - there are frustratingly few references in recent literature that focus on bona fide spectroscopic EUV observations of the lower solar atmosphere during flares. (See Table~\ref{tab:eis_fp_flares} for a list of the 18 flares successfully observed by EIS during the impulsive phase to date in the 9 years since Hinode was launched, along with the associated publications in which they appear.) This is perhaps largely in part due to the difficulties associated with positioning the slit from a rastering spectrometer with a reduced field of view precisely over the footpoints or ribbons during the crucial few minutes of an impulsive phase, or detecting weak chromospheric enhancements against the strong EUV background in full-disk irradiance measurements. A reduced duty cycle is a factor for instruments such as Thermosphere Ionosphere Mesosphere Energetics Dynamics/Solar EUV Experiment (TIMED/SEE; \citealt{wood04}) that only made measurements a few times per day, as well as spectrographs such as SERTS (Solar EUV Rocket Telescope and Spectrograph) and EUNIS (EUV Normal Incidence Spectrographs) that flew on sounding rockets and only observed the Sun for a few minutes at a time. The few references that are available, however, have gone a long way in increasing our understanding of the dynamic response of the lower solar atmosphere during flares. In doing so, these studies have also highlighted the gaps in our knowledge, and pointed to the questions that can yet be answered by spectroscopic EUV observations, particularly when used in conjunction with those from other instruments at different wavelengths, as well as theoretical modelling. This review paper aims to summarise many of the significant findings that have arisen from spectroscopic observations of flare footpoints and ribbons over the past decade or so in addition to raising awareness of the issues that remain to be addressed.

\begin{center}
\begin{table}[!t]
   \begin{adjustbox}{max width=\textwidth}
   \caption{Flares Whose Footpoints Were Observed by Hinode/EIS During the Impulsive Phase and the publications in which they appear}
   \begin{threeparttable}
   \begin{tabular}{lcc}
   \hline
					&GOES	 		&						\\
Flare ID				&Class			&Associated Publications		\\
   \hline
SOL2007-01-16T02:22	&C4.1			&\cite{wata09,li11}			\\
SOL2007-05-22T23:26	&B2.7 			&\cite{delz11,grah13}		\\
SOL2007-06-02T20:32	&B2.7\tnote{a,b,c}	&\cite{delz08,warr08}		\\
SOL2007-06-05T15:50	&C6.6			&\cite{grah11}				\\
SOL2007-06-06T16:55	&C9.7			&\cite{wata10}				\\
SOL2007-06-07T00:37	&B7.6\tnote{b}		&\cite{mill08}				\\ 
SOL2007-12-07T04:35	&B1.4			&\cite{chen10}				\\
SOL2007-12-14T15:22	&B8.8 			&\cite{grah13}				\\
SOL2007-12-14T15:54	&B8.8 			&\cite{grah13}				\\
SOL2007-12-14T01:39	&B9.6 			&\cite{grah13,grah14}		\\
SOL2007-12-14T14:16	&C1.1			&\cite{mill09,mill11},			\\
					&				&\cite{ning11,grah13},		\\
					&				&\cite{grah14}				\\
SOL2007-12-16T06:23	&B1.8 			&\cite{grah13}				\\
SOL2011-02-13T13:44	&C4.7			&\cite{li14}				\\
SOL2011-02-16T01:32	&M1.0			&\cite{li12}				\\
SOL2011-02-16T07:35	&M1.1			&\cite{youn13}				\\
SOL2012-03-07T18:49	&C1.4\tnote{c}		&\cite{bros13a}				\\
SOL2012-03-09T03:22	&M6.3\tnote{a}		&\cite{dosc13}				\\
SOL2012-11-21T18:20	&C1.9\tnote{c}		&\cite{bros13b}				\\
   \hline
   \end{tabular}
   \begin{tablenotes}
   \item[a]{Both full CCDs were read out for these events}
   \item[b]{Footpoints were observed but the flare was not impulsive (i.e. there were no associated hard X-rays)}
   \item[c]{Not included in the GOES Event List}
   \end{tablenotes}
   \end{threeparttable}
   \label{tab:eis_fp_flares}
   \end{adjustbox}
\end{table} 
\end{center}

\section{Historical Observations}
\label{s:history}

Impulsive EUV bursts associated with a flare's impulsive phase were first discovered in the late 1960's and early 1970's using (photometric) data from the Orbiting Solar Observatory (OSO) III satellite \citep{hall69,hall71}. These bursts were primarily observed in cool emission lines, such as He~II, C~II, O~V, and Si~III. Improved observations by the Harvard College Observatory (HCO) EUV Spectroheliometer on Skylab showed similar bursts at even higher temperatures (e.g., Mg~X; \citealt{emsl78a}). Around the same time, increases in EUV emission were also being inferred from Sudden Frequency Deviations (a subset of Sudden Ionospheric Disturbances) observations by \cite{donn69}, \cite{kane71}, and \cite{donn78}. While these observations were not spectroscopic per se, they were found to be temporally correlated with HXR emission. This suggested that the EUV emission was predominantly chromospheric in nature given that the ionosphere is sensitive to emission in the 10--1030\AA\ range, which is dominated by chromospheric free-bound and bound-bound transitions \citep{emsl78b}.

Observations of the Lyman continuum (LyC) of neutral hydrogen from the quiet Sun detected by OSO-IV (and later OSO-VI) were first reported by \cite{noye70} and \cite{vern72}. These data were used to determine the structure of the chromosphere, including temperature, density, and opacity, using the Eddington-Barbier relationship. \cite{mach78} later applied this same technique to a number of solar flares observed using HCO to reveal that the ground level of hydrogen is closer to local thermodynamic equilibrium during flares than in quiescent times, and that the layer at which LyC is formed during flares is lower in the atmosphere. This was verified by later studies (e.g. \citealt{mach80}), including what appears to be the only reference in the literature of imaging a flare in the Lyman continuum \citep{hira85}. This was carried out using the Japanese sounding rocket S520-5CN. These findings were also used to determine the mass of the material evaporated into the overlying loops. As LyC is formed in the upper chromosphere/lower transition region it is more sensitive to heating from above compared to the Balmer and Paschen continua, which are formed in deeper layers. While this feature of Lyman continuum emission points to an important flare diagnostic, few other reports of continuum enhancements exist in the literature. Despite this lack of observations, \cite{ding97} improved on the method of \cite{mach78} to include stimulation by nonthermal electrons in anticipation of future observations of these vital emissions. Similar measurements of the Lyman continua of He~I \citep{ziri75} and He~II \citep{lins76} were also carried out in the 1970's.

\section{EUV Diagnostics}
\label{s:euv_diag}

\begin{figure}[!t]
\centerline{\includegraphics[width=\textwidth]{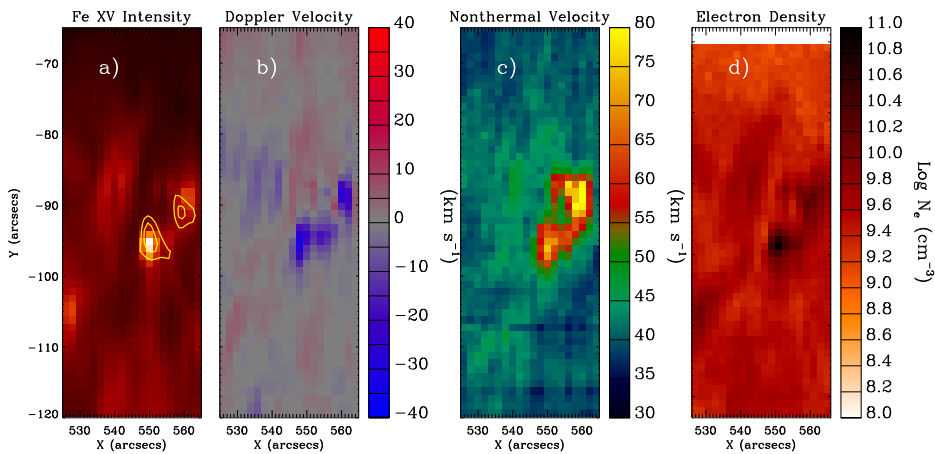}}
\caption{Derived plasma parameters from a single Hinode/EIS raster taken during the impulsive phase of a C1.1 flare that occurred on 14 December 2007. a) An image showing the spatial distribution of the Fe~XV 284\AA (2~MK) line intensity. Two bright footpoints of an overlying flare loop are visible near the middle of the raster. Overlaid are the contours of the 20-25~keV sources as observed by RHESSI that denote the location of the HXR footpoints. b) The corresponding Doppler velocity map derived from shifts in the line centroid relative to a quiet-Sun value. Positive velocities (redshifts) indicate downflows, while negative velocities (blueshifts) indicate upflows. c) Map of the nonthermal velocity from the line widths over and above the thermal-plus-instrumental widths. d) Spatial distribution of electron density from the ratio of two Fe~XIV lines which are formed at a similar temperature to that of Fe~XV. This figure shows that plasma at the location of the HXR emission (believed to be in the chromosphere) was hot, upflowing, turbulent, and dense. From \cite{mill11}.}
\label{eis_fp_rasters}
\end{figure}

The introduction of imaging spectrometers such CDS, and its successor EIS, allowed spatially-resolved line profiles to be obtained by stepping a slit across a chosen section of the solar atmosphere, building up a three-dimensional raster ($x(t)$, $y$, $\lambda$). Exposure (and read-out) times were initially quite long ($\gtrsim$30s), and so depending upon the chosen field-of-view, a single raster could often take as much as an hour or more to complete. EIS improved upon this with its greater sensitivity and broader temperature coverage, and rasters featuring a breadth of emission lines could be run in minutes making it more ideal for flare studies. For example, on 14 December 2007 EIS observed a simple, isolated C1.1 flare while running a fast-raster (3.5~minute cadence) study that contained over 40 individual emission lines covering the 10$^{4}$--10$^{7}$~K temperature range, including several density sensitive line pairs. Figure~\ref{eis_fp_rasters}$a$ shows the raster taken during the flare's impulsive phase, clearly showing the two footpoints (in Fe~XV in this case) that coincided with the 20--25~keV hard X-ray emission observing by the Ramaty High Energy Solar Spectroscopic Imager (RHESSI; \citealt{lin02} - orange contours). Figures~\ref{eis_fp_rasters}$b$, $c$, and $d$ show the corresponding Doppler velocity, nonthermal velocity, and density maps (from Fe~XIV line ratios), respectively. The data obtained from this raster showed that material at the footpoints of this flare was hot, evaporating, turbulent, and dense, and EIS is able to do this for multiple emission lines simultaneously. However, by January 2008 Hinode had lost the use of its X-band transmitter and so the available telemetry was greatly reduced. The findings that resulted from this and other, similar studies are discussed in sections~\ref{ss:lines_int}, \ref{ss:lines_vel}, \ref{ss:lines_wid}, and \ref{ss:lines_rat}, while section~\ref{ss:continuum} deals with recombination continuum observations from EVE.

Many of the diagnostic methods discussed in the following subsections implicitly assume that the plasma under investigation is in ionisation equilibrium. This assumption may be valid at high densities when ionization timescales are short \citep{brad09,grah13,casp14}, but at lower values departures from equilibrium can be significant, particularly for cooler emission lines \citep{doyl13}. Such departures have been found during impulsive heating of cool, diffuse, coronal plasma \citep{brad06} and in sunspot plumes \citep{doyl85}. The consequence of non-equilibrium ionisation during flare heating would be that emission lines under investigation could be formed at higher or lower temperatures than otherwise assumed. The problem can be exacerbated for plasma in motion. For material moving upwards (or downwards) into regions of higher (or lower) temperatures, recombination rates can be out of sync with ionisation. For example, \cite{fran81} found the the formation temperature of the C~IV resonance lines dropped from log~T=5.0 to log~T=4.65 for a downflow of 6 km~s$^{-1}$. The integration times of the observations under investigation will also play a role. For long ($>$tens of seconds) exposure times, short-lived departures from equilibrium may well be averaged out. However, as instrumentation becomes more advanced, and observational timescales become shorter, departures from equilibrium may becomes significant enough to bias the interpretation (see Section~\ref{s:future}).

\begin{figure}[!b]
\begin{center}
\includegraphics[width=0.48\textwidth]{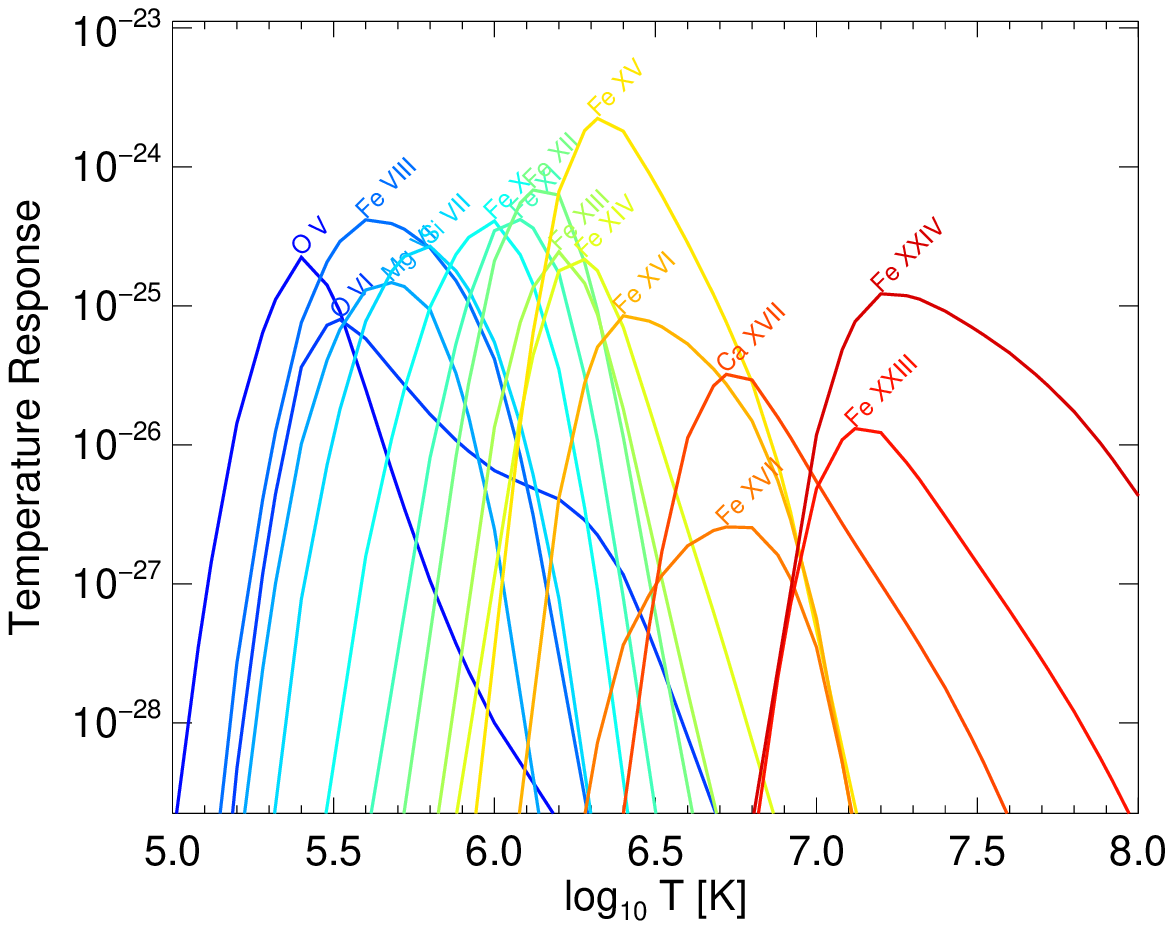}
\includegraphics[width=0.48\textwidth]{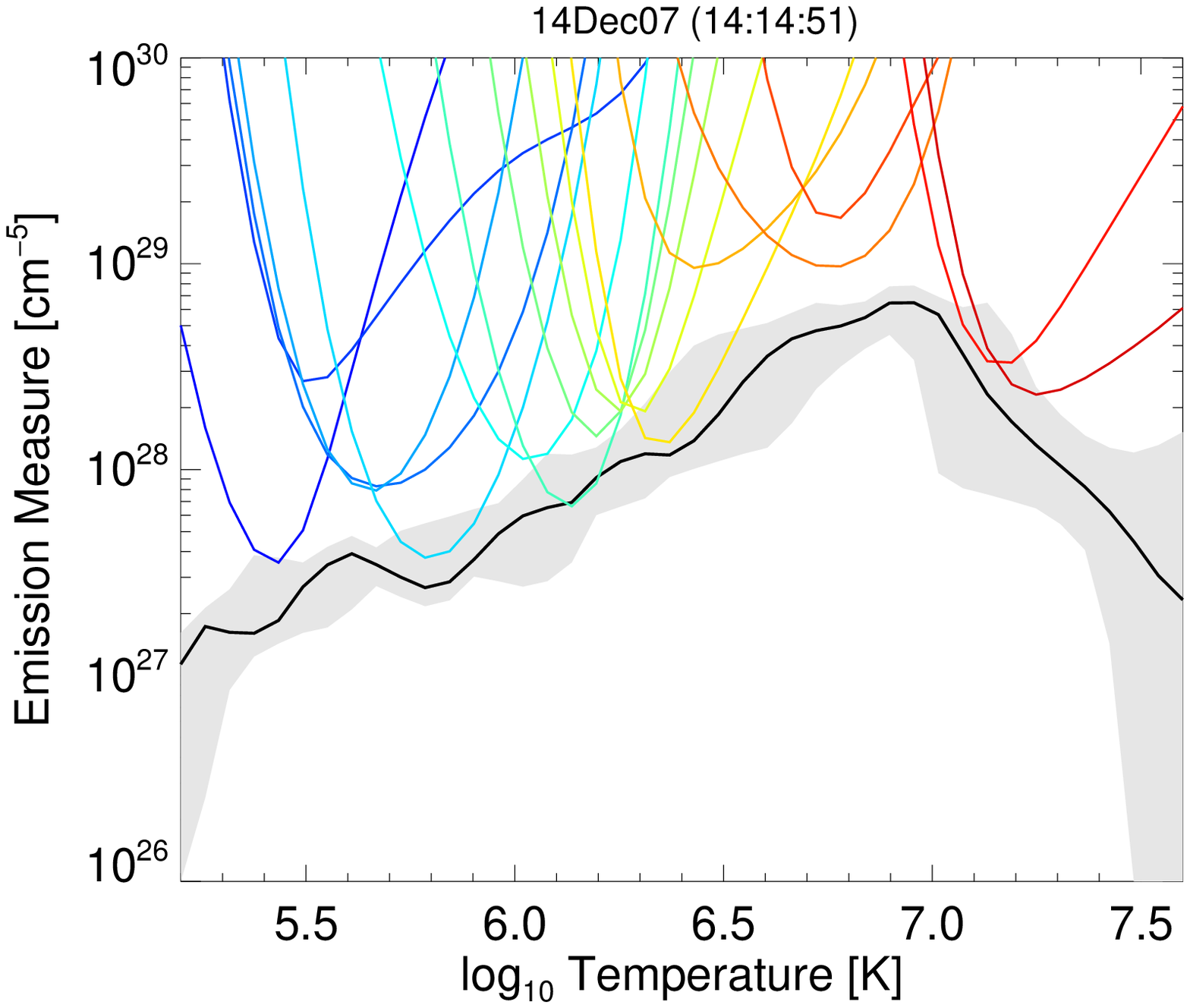}
\caption{Left: Contribution functions for 16 lines observed by Hinode/EIS. Right: A DEM for a flare footpoint during the impulsive phase of the 14 December 2007 C-class flare. Grey shaded area denotes the corresponding error bars. From \cite{grah13}.}
\label{fp_dem}
\end{center}
\end{figure}

\subsection{Emission Line Intensities} 
\label{ss:lines_int}
The intensity of emission lines formed at discreet temperatures can be used to reconstruct a differential emission measure (DEM) profile of the emitting plasma. The DEM is a powerful tool useful for investigating the temperature distribution of the emitting plasma, providing crucial clues to the underlying heating processes. Despite decades of study, there is still no consensus on what the flare DEM profile looks like (e.g., \citealt{mcti99}), particularly during the impulsive phase. However significant advances are currently being made using EVE data, with RHESSI data being used to constrain the highest temperature emission \citep{warr13,casp14}. For example, the often-assumed flare DEM used in the CHIANTI atomic database is based on a single observation by \cite{dere79} taken during the decay phase of an M-class flare. However, it is now possible to determine the DEM from a flaring footpoint using spatially-resolved line profiles. The right-hand panel of Figure~\ref{fp_dem} shows a regularised DEM solution from \cite{grah13} of a spatially-resolved footpoint from the impulsive phase of the 14 December 2007 flare using data from EIS using the methods of \cite{hann12}. The colored lines are the EM loci curves (ratio of the data and the temperature response) that define the isothermal (and hence maximum) emission possible at a given temperature. \cite{grah13} also used this technique on a number of B-class flares to establish the distribution of temperature at the footpoints. They found the slope of the distribution between log~T=5.5--7.0 in each case to be $\sim$1 (EM(T)$\propto$T), which they concluded was an indication that the layers beneath the electron deposition site were being heated conductively. 

\begin{figure}[!t]
\begin{center}
\includegraphics[width=0.49\textwidth]{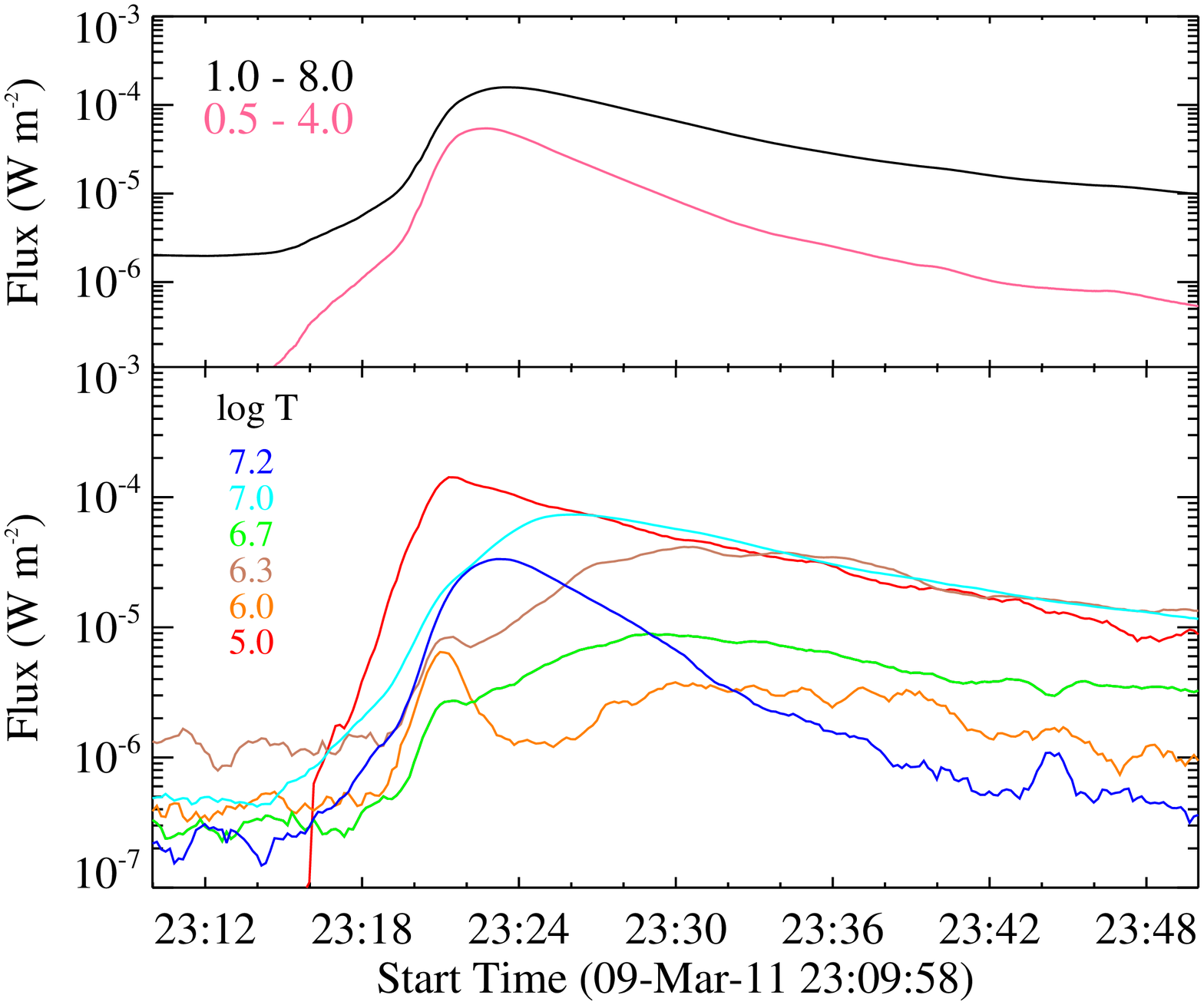}
\includegraphics[width=0.49\textwidth]{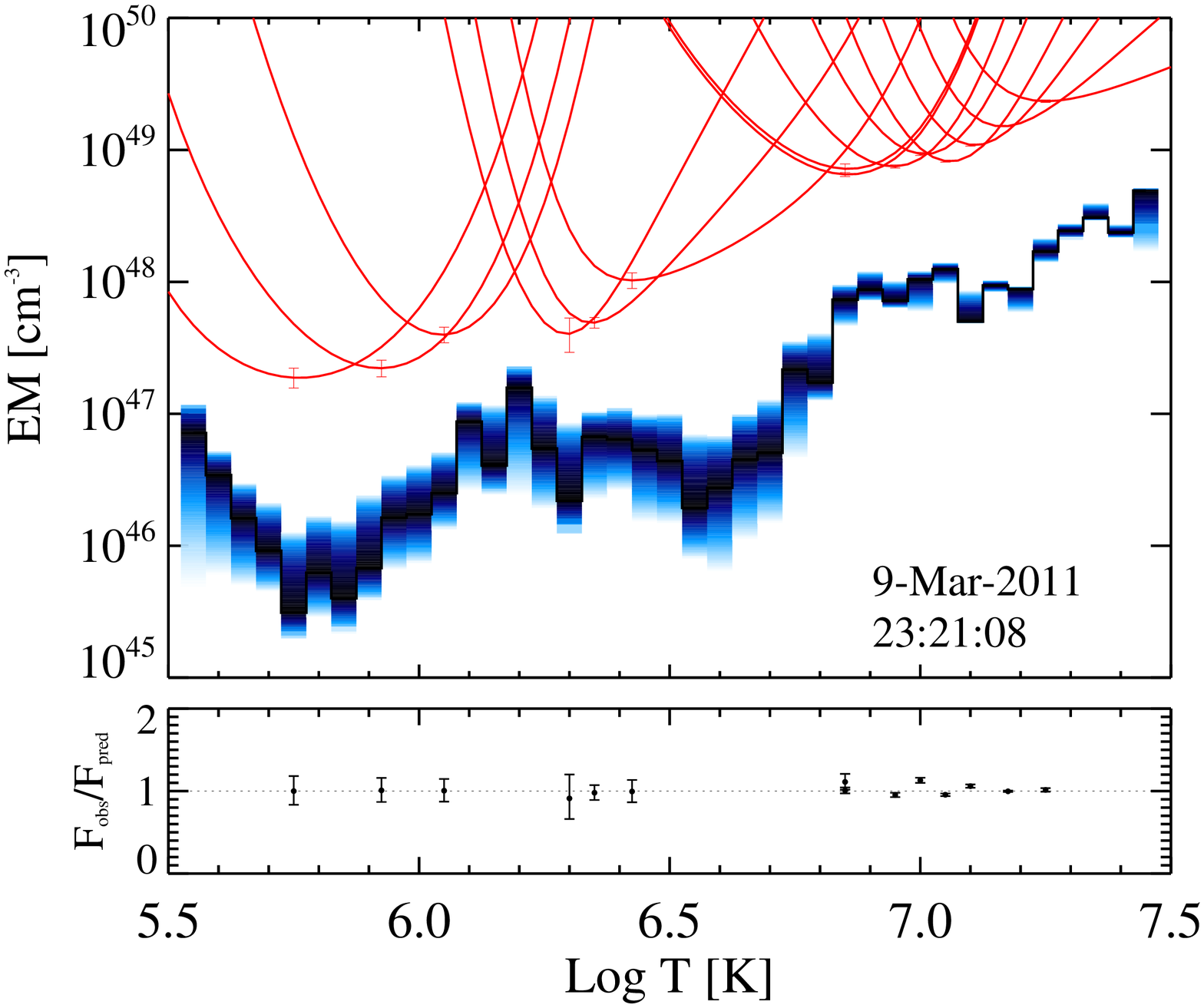}
\includegraphics[width=0.49\textwidth]{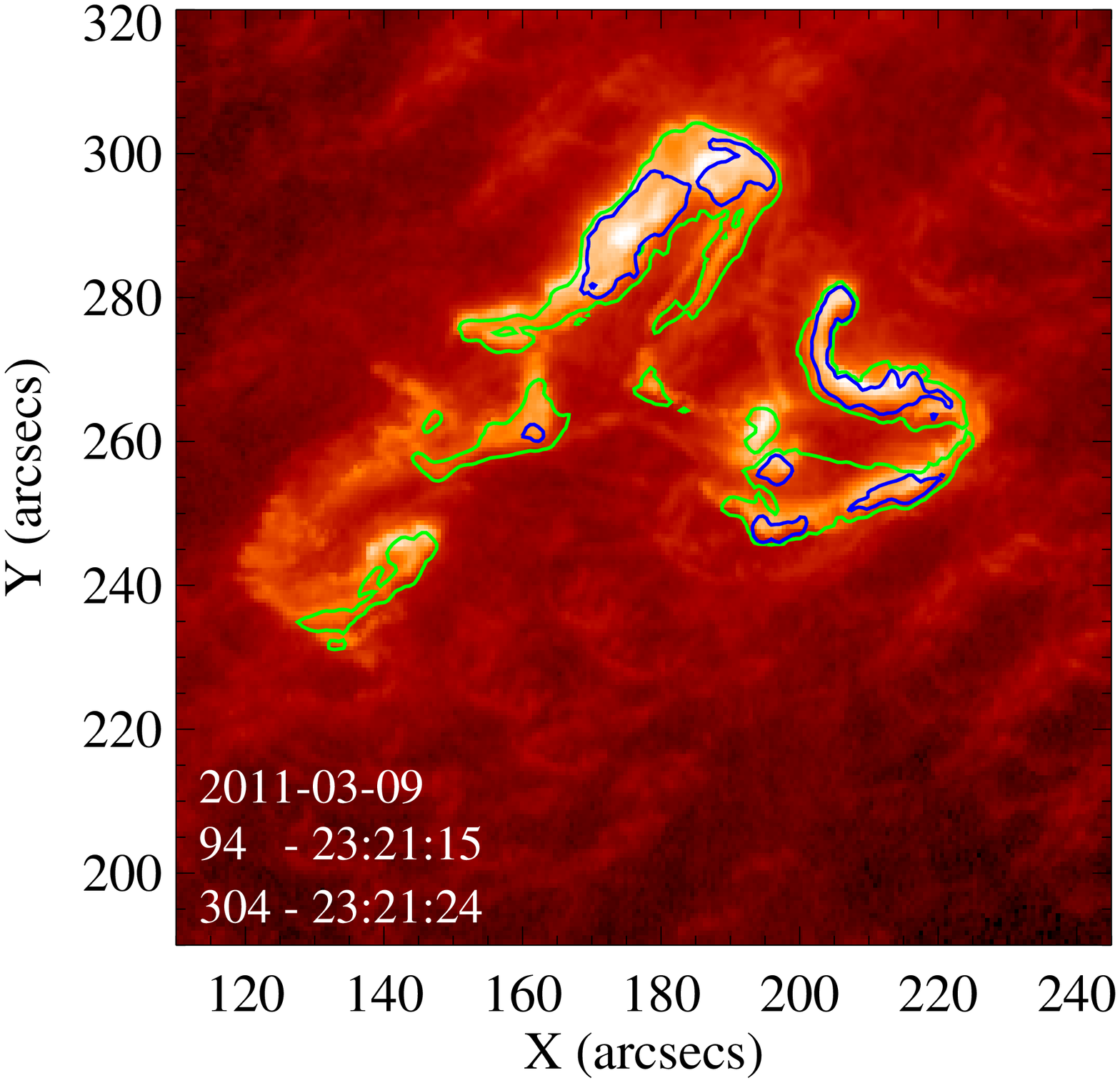}
\caption{Top left panel: lightcurves of SXR emission (top panel) and EUV emission binned by temperature (bottom panel) for an M1.1 flare that occurred on 9 March 2011. Note that the log~T = 6.0, 6.3, and 6.7 curves peak both during the impulsive phase as the chromosphere is heated - in sync with He~II - and again during decay phase as the flare loops cool. Top right panel: DEM derived from emission lines observed by EVE during the impulsive phase (23:21~UT). Bottom panel: SDO/AIA 304\AA\ image also taken during the impulsive phase. Overlaid are the contours of the associated 94\AA\ emission, which is dominated by Fe~XVIII emission (log~T=6.9), and which also aligns with the He~II ribbons. From \cite{kenn13}.}
\label{eve_fp_dem}
\end{center}
\end{figure}

By contrast, looking at larger ($>$M1) events observed by EVE, \cite{kenn13} found a double-peaked DEM (top right hand panel of Figure~\ref{eve_fp_dem}). Although EVE provides spatially-integrated (Sun-as-a-star) observations, comparison between lightcurves binned by temperature (top left hand panel of Figure~\ref{eve_fp_dem}; see \citealt{cham12}) and associated AIA images revealed emission with temperatures approaching 8~MK to be both co-temporal and co-spatial with He~II 304\AA\ ribbons during the impulsive phase (bottom panel of Figure~\ref{eve_fp_dem}). This implied that EVE spectra taken during this time, and therefore the resultant DEM, would be indicative of predominantly footpoint emission. The reason for the different DEM profiles obtained by \cite{grah13} and \cite{kenn13} is unclear, although it may simply be attributed to the two studies focusing on events of very different magnitudes. The nature of the observations may also play a role. While EIS provides spatially resolved line profiles, they are often only for a single time interval when the slit of the spectrometer happened to raster over the footpoint. EVE, on the other hand, can provide continuous spectra throughout a flare, but with no spatial information. Simultaneous flare observations acquired with both instruments would greatly help in resolving this issue.

\subsection{Emission Line Shifts} 
\label{ss:lines_vel}
In the context of solar flares, SXR spectroscopy led to the first quantiative measurements of mass flows due to chromospheric evaporation through Doppler shifts of high-temperature line profiles \citep{dosc80,anto83,canf87,dosc92,dosc96}. However, these studies often focused on a single, high-temperature line and lacked any spatial information. Line profiles were found to be dominated by a strong stationary component with a blue-wing asymmetry indicative of upflows of several hundred km~s$^{-1}$. It was initially thought that the stationary component  emanated from the loop tops when flows had ceased or were moving perpendicular to the line of sight, while the blue wing of the line was an indicator of evaporating material at the footpoints (see \citealt{dosc05}). The introduction of CDS on SOHO enabled spatially-resolved line profiles to be measured across a range of temperatures, although the sampling was still quite sparse \citep{czay99,teri03,mill06a,mill06b,delz06,raft09}. Despite the provision of spatially-resolved line profiles from CDS, the lines still appeared to be dominated by a stationary component in contrast to the predictions of theory \citep[e.g.][]{li89}. The use of sit-and-stare observations further allowed flow velocities to be determined as a function of time but at the cost of large uncertainties in the pointing aspect \citep{bros03,bros04,bros07,bros09a,bros09b}.
 
\begin{figure}[!t]
\begin{center}
\includegraphics[height=\textwidth,angle=90]{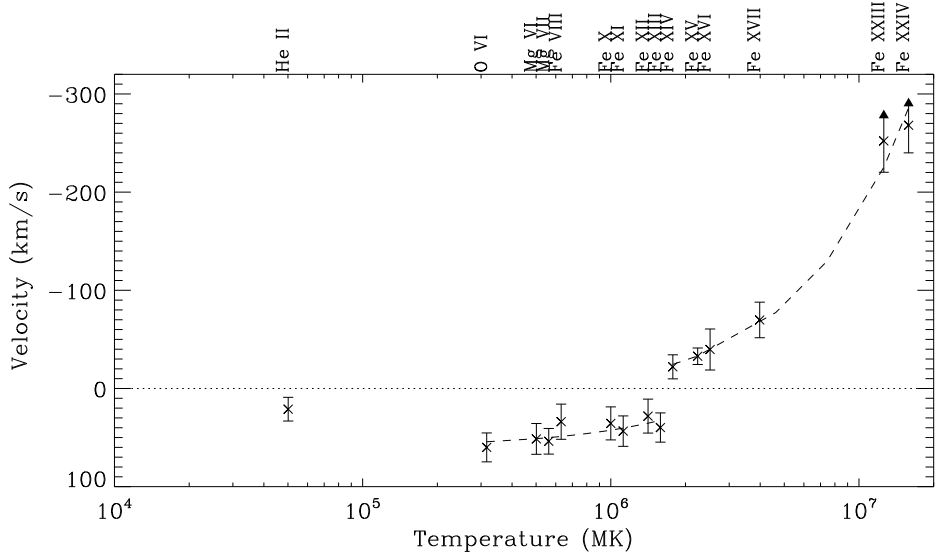}
\caption{Derived Doppler velocities at a flaring footpoint for 15 emission lines formed between 0.05--16~MK. Negative velocities (blueshifts) represent upflows while positive velocities (redshifted) denote downflows. From \cite{mill09}.}
\label{eis_vel_temp}
\end{center}
\end{figure}

EIS now allows spatially-resolved line profiles to be measured using emission lines formed over a broad range of temperatures simultaneously \citep{mill09,delz11,grah11}. Figure~\ref{eis_vel_temp} shows the derived Doppler velocities for 15 emission lines, formed between 0.05--16~MK, taken from one of the footpoints of the 14 December 2007 flare \citep{mill09}. This shows a linear relationship between the temperature of the heated plasma and the rate at which it rises ($v_{up} \propto T$). The material that was redshifted (chromospheric condensation, driven downward by the overpressure of the evaporated material required to conserve momentum; \citealt{canf87,canf90}), recoiled at approximately the same speed regardless of temperature but was also observed at much higher temperatures than previously expected from chromospheric evaporation models (up to 1.5~MK). \cite{fish85} believed that the temperature at which the Doppler velocity changed from redshifted to blueshifted (dubbed the `Flow Reversal Point' by \citealt{bran14}) marked the layer of the solar atmosphere at which the electrons deposited the bulk of their energy. From \cite{mill09}, this implied that either the electrons lost their energy in the corona (unlikely, due to the low densities involved) or that the chromosphere was being heated as it recoiled. More recent simulations now show that hot redshifted material is possible if chromospheric heating is sustained \citep[e.g.][]{liu09}, rather than by a single burst as assumed by \cite{fish85}. Similar findings of hot redshifts have also been made by \cite{mill08}, \cite{chen10}, and \cite{grah11}, although the precise `crossover' temperature varied between 0.5 and 2~MK for different events. In one case presented by \cite{li11} it was found to be as high as 5~MK.

Despite its modest spectral resolution, \cite{huds11} were able to use EVE data to detect redshifted He~II emission associated with chromospheric condensation during flares. By measuring Doppler velocities during the impulsive phase relative to those in the decay phase, they found downflows of 10--20~km~s$^{-1}$, consistent with those from EIS observations. Diminishing blueshifts in Fe~XXIV, as expected from chromospheric evaporation theory, were also detected throughout the main phase of the same events, although, as with the He~II velocities, the authors correctly advised caution when attempting to derive the absolute magnitude of the flows. Many lines in the EVE data are likely to be significantly blended, and the lack of spatial information introduces line-of-sight effects due to flare location and loop geometry.

The issue of whether the stationary or blueshifted component should dominate the high-temperature ($\gtrsim$10~MK) line profiles is still an open question, with different studies promoting either case: \cite{mill09} and \cite{li11} both found stationary emission to dominate, while \cite{wata10} and \cite{youn13} found lines to be completely blueshifted. In one particularly complex flare, \cite{dosc13} found both scenarios to occur at different locations along the ribbons. It may be the case that current instrumentation does not have high enough spatial resolution to resolve the evaporation flows as they commence before they become `diluted' with coronal material. New observations of Fe~XXI emission (1354.1\AA) from the recently-launched Interface Region Imaging Spectrograph (IRIS; \citealt{pont14}) may help address this issue \citep[see][]{youn15}.

\subsection{Emission Line Widths}
\label{ss:lines_wid}
The width of spectral emission lines contains important information on the temperature and turbulence of the emitting plasma. Line width is generally made up of at least three components: the intrinsic instrumental resolution, the thermal Doppler width, and any excess (nonthermal) broadening, which may be an indicator of possible turbulence, or pressure or opacity broadening. Many studies have reported excess SXR line broadening over and above that expected from thermal emission during a flare's impulsive phase indicating possible turbulent motion \citep{dosc80,feld80,gabr81,anto82} although this emission was integrated over the entire solar disk. 

\begin{figure}[!t]
\centerline{\includegraphics[height=\textwidth,angle=90]{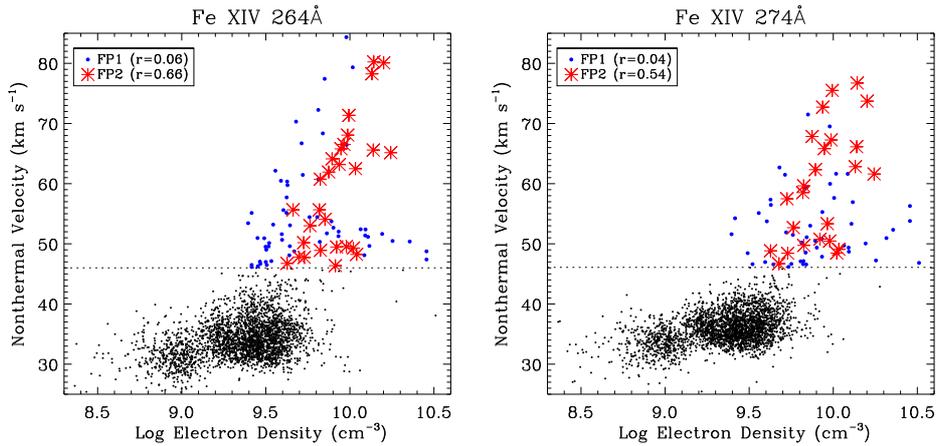}}
\caption{Plot of nonthermal velocity derived from the width of two Fe~XIV lines (264\AA; left, 274\AA; right) against electron density derived from the ratio of two Fe~XIV lines during the impulsive phase of the 14 December 2007 flare. Black dots denote the surrounding quiescent active region while blue circles and red asterisks are data points from the two footpoints. A clear correlation exists between the two parameters for at least one of the footpoints (FP2).}
\label{eis_nthvel_den}
\end{figure}

\cite{mill11} presented a detailed investigation into the nature of spatially-resolved line broadening of EUV emission lines during the impulsive phase of the 14 December 2007 flare using EIS thanks to its superior spectral and spatial resolution. Line profiles, co-spatial with HXR emission observed by RHESSI, were found to be broadened beyond their thermal widths (Figure~\ref{eis_fp_rasters}$c$). Using techniques similar to those used to establish the nature of line broadening in quiescent active region spectra \citep{hara08,dosc08,brya10,pete10}, it was found that a strong correlation existed between Doppler velocity and nonthermal velocity for the Fe~XV and Fe~XVI lines at one of the footpoints. This suggested that the line broadening at these temperatures was a signature of unresolved plasma flows along the line of sight during the process of chromospheric evaporation. Analysis of Fe~XIV lines, on the other hand, which showed no conclusive correlation between Doppler and nonthermal velocities, showed a stronger correlation between electron density and nonthermal width (Figure~\ref{eis_nthvel_den}) suggesting that the excess line broadening at these temperatures may have been due to opacity or pressure broadening as emitted photons were scattered along the line of sight. However, estimating the opacity of the emission, $\tau_{0}$, using:
\begin{equation}
\tau_{0} = 1.16 \times 10^{-14} \lambda f_{ij} \sqrt{\frac{M}{T}} \frac{n_{ion}}{n_{el}} \frac{n_{el}}{n_{H}} \frac{n_{H}}{N_{e}} N_{e}h
\label{opacity_eqn}
\end{equation}

\noindent
where $\lambda$ is the wavelength of the line, $f_{ij}$ is the oscillator strength, $M$ is the mass of the ion, $n_{ion}/n_{el}$ is the ionisation fraction, $n_{el}/n_{H}$ is the abundance of iron relative to hydrogen, $N_{e}$ is the electron density, and $h$ is the column depth, and the amount of pressure broadening, $\Delta \lambda$, using:
\begin{equation}
\Delta \lambda = \frac{\lambda^{2}}{c} \frac{1}{\pi \Delta t_{0}} \approx \frac{\lambda^{2}}{c} \frac{N_{e} \sigma}{\pi} \sqrt{\frac{2k_{B}T}{M}}
\label{pressure_eqn}
\end{equation}

\noindent
where $\Delta t_{0}$ is the timescale of the emitting photon and $\sigma$ is the collisional excitation cross-sectional area (from \citealt{bloo02} and \citealt{chri06}), suggested that the amount of excess broadening should have been negligible in each case. Perhaps the assumptions made in solving Equations~\ref{opacity_eqn} and \ref{pressure_eqn} were incorrect (e.g. lines were not formed in ionization equilibrium), or the broadening was due to a culmination of different effects, or perhaps it was due to a different mechanism altogether not considered (e.g. Stark broadening). Perhaps larger, more energetic events, or density diagnostics of higher (or lower) temperature plasmas, will show these effects to be more substantial. \cite{bros13b} found nonthermal velocities derived from the width of Mg~VI, Si~VII, Fe~XIV, Fe~XVI, and Fe~XXIII emission lines increased by a factor of $\sim$3 compared to quiet-Sun widths at the site of a flare ribbon using sit-and-stare observations. This was assumed to be an indicator of turbulent motion associated with chromospheric evaporation. Line broadening can not only reveal important information with regard to the heating processes during flares but can also be a crucial diagnostic of the fundamental atomic physics (abundances, ionisation fractions, etc.) and must therefore be a component of future flare modelling.

\subsection{Emission Line Ratios} 
\label{ss:lines_rat}

Values of the electron density of a given plasma can be found by taking the ratio of the flux of two emission lines from the same ionization stage when one of the lines is derived from a metastable transition (see Chapter 2.8.2 of \citealt{mari92} for a detailed derivation of this technique and Chapter 4.4 for early applications to solar observations). This approach is more robust than using broadband filter ratios (e.g. \citealt{whit05}) that require an accurate knowledge of the volume of the emitting plasma, which can be difficult for plane-of-sky observations. A filling factor of unity is also often assumed. Line ratio techniques have more recently been employed to infer electron densities in active regions \citep{gall01,mill05,dosc07}, jets \citep{chif08}, and flare loops \citep{mill12b}.

\begin{figure}[!b]
\begin{center}
\includegraphics[height=0.49\textwidth,angle=90]{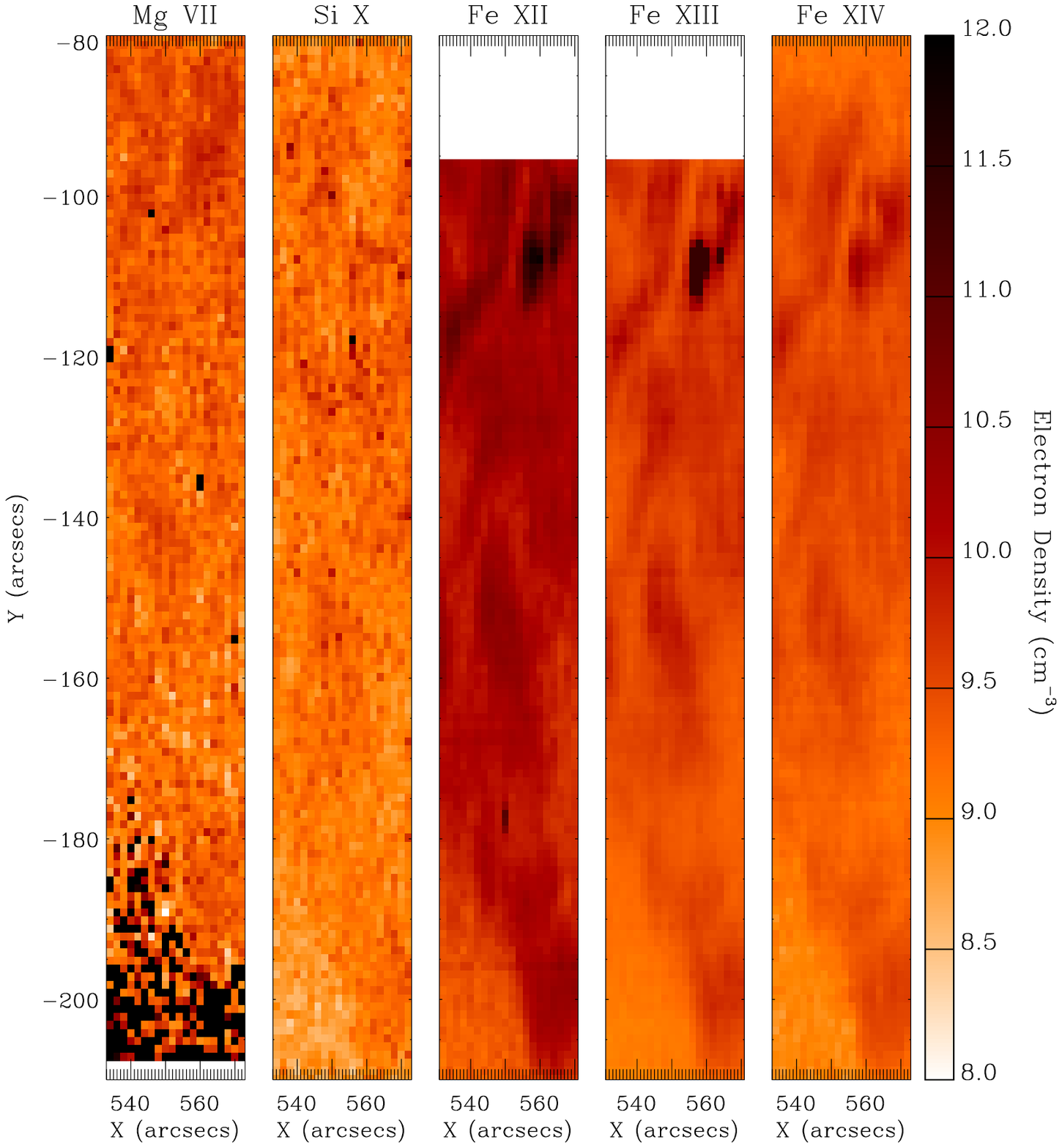}
\includegraphics[height=0.49\textwidth,angle=90]{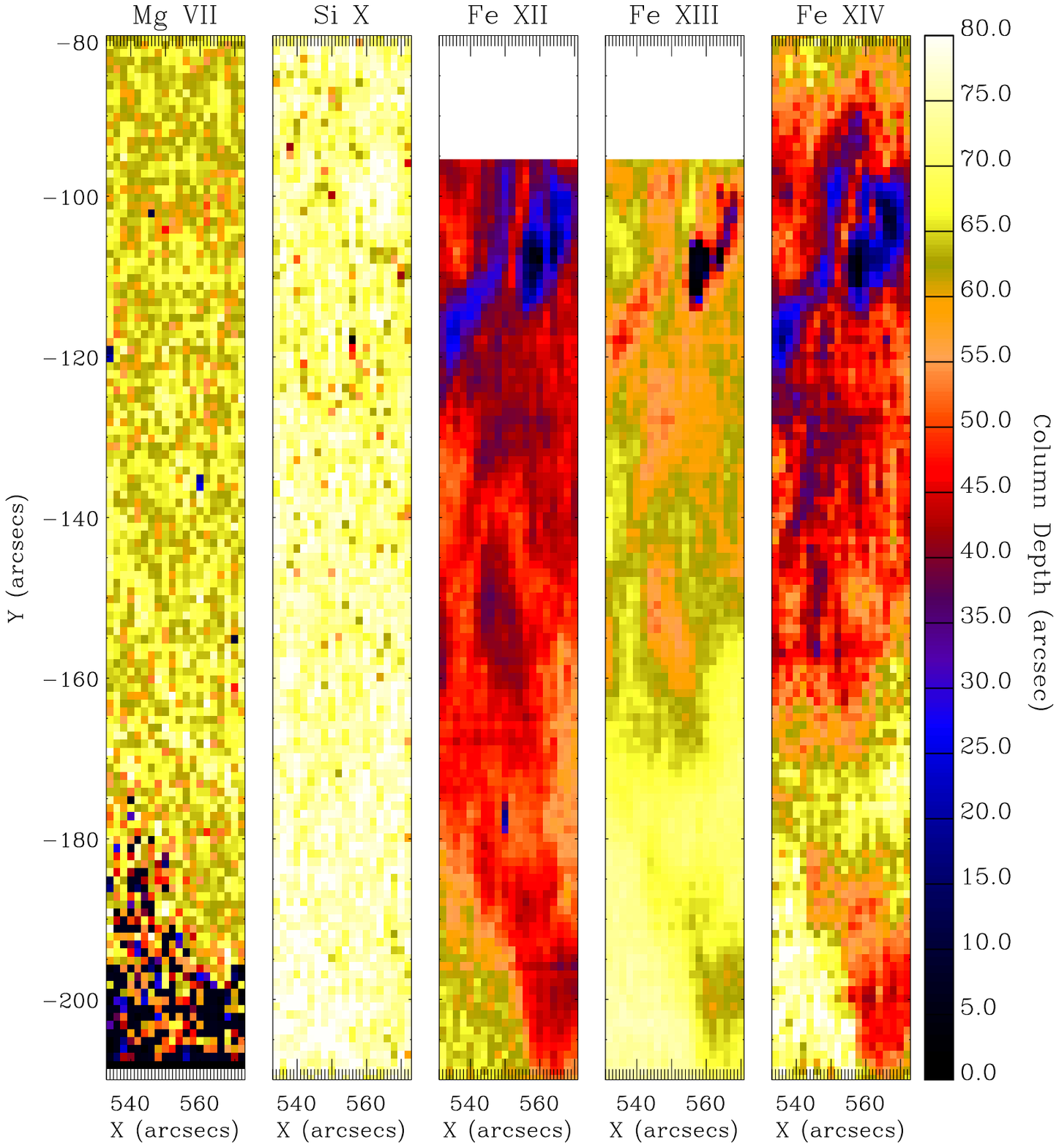}
\includegraphics[height=0.49\textwidth,angle=90]{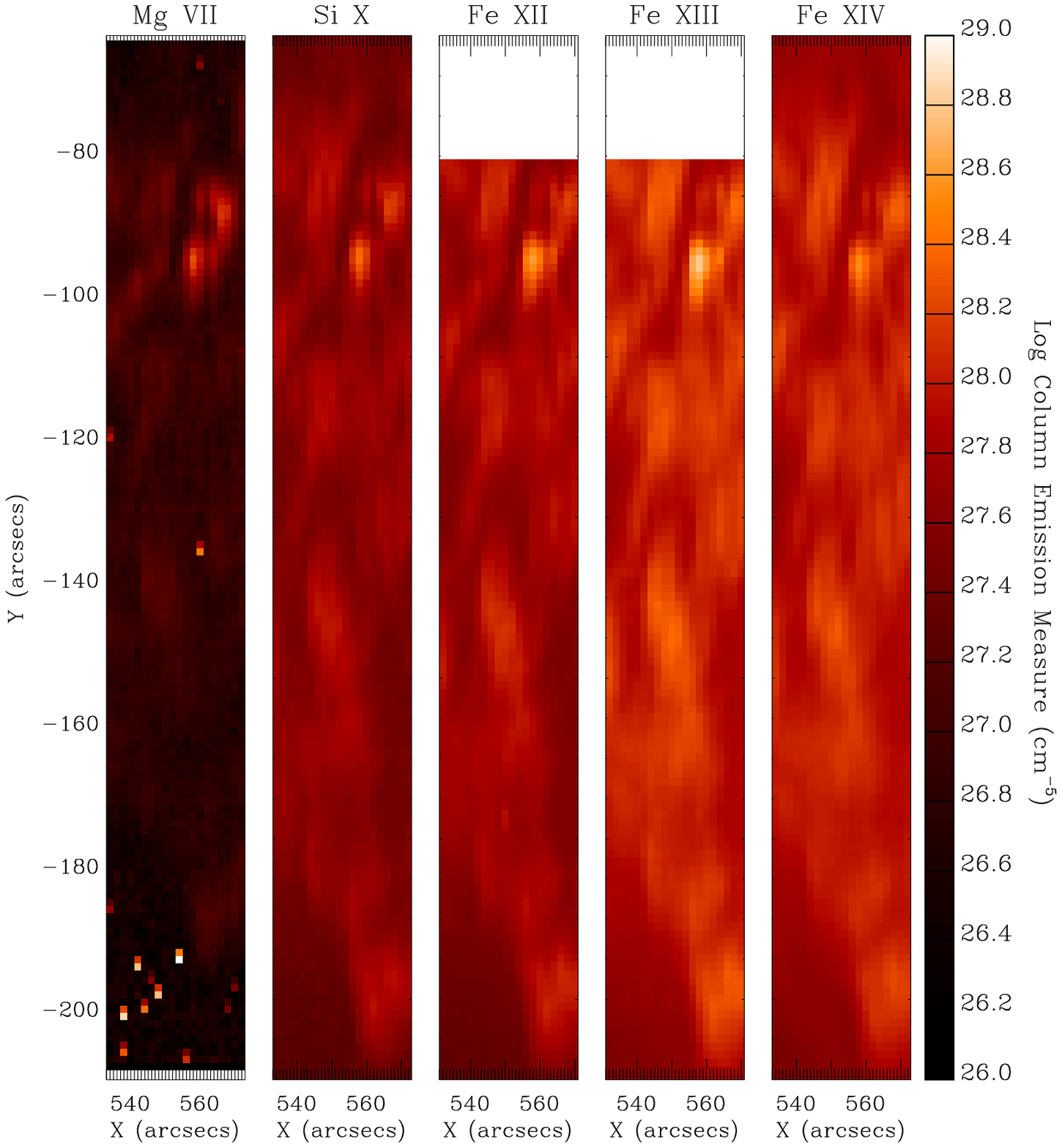}
\caption{Five EIS rasters in the Mg~VII, Si~X, Fe~XII, Fe~XIII, and Fe~XIV lines, taken during the impulsive phase of the 14 December 2007 flare. Top left: Electron density maps formed using line ratio techniques. Top right: Maps of column depth formed using Equation~\ref{col_depth_two}. Both figures are taken from \cite{mill11}. Bottom: Column emission measure maps generated by combining data from the top two panels. Taken from \cite{grah13}.}
\label{eis_den_col_dep_maps}
\end{center}
\end{figure}

\cite{mill11} used five pairs of density sensitive line ratios formed at different temperatures (Mg~VII, Si~X, Fe~XII, Fe~XIII, and Fe~XIV) to measure density enhancements at the footpoints during the 14 December 2007 flare. Peak values of 10$^{11}$--10$^{12}$~cm$^{-3}$ were found using Fe~XII and Fe~XIII ratios, indicating enhancements of 2--3 orders of magnitude above quiescent active region levels. It was found that density enhancements were only visible in lines formed above $\sim$1~MK as cooler lines are formed deeper in the solar atmosphere where densities are already high (Figure~\ref{eis_fp_rasters}$d$ and top left hand panels in Figure~\ref{eis_den_col_dep_maps}). A recent follow-up study by \cite{grah14} found density values approaching 10$^{12}$~cm$^{-3}$ using O~V lines for the same event. \cite{grah11} also found comparable values from Fe~XII, Fe~XIII, and Fe~XIV ratios at the location of the ribbons during a C6.6 flare. Sit-and-stare EIS observations from \cite{bros13b} also revealed a rapid increase in electron density of an order of magnitude using an Fe~XIV line pair, but this increase appeared to be independent of the corresponding upflow velocity as would be expected in a beam-heated chromosphere scenario; the density (and intensity) enhancements ceased while blueshifts persisted, which the author stipulated as evidence for an energy release process in the chromosphere itself as opposed to heating from above. However, care must be taken when interpreting observations from single slit measurements as pointing accuracy and co-alignment with context images can be ambiguous.

Using the values derived for the electron densities, it is possible to compute the column depth, $h$, of the emitting material at the formation temperature of the lines. The intensity of a given emission line, $I$, can be expressed as:

\begin{equation}
4 \pi I = 0.83 \int G(T, N_{e}) N_{e}^{2}dh
\label{col_depth_one}
\end{equation}

\noindent
where $G(T,N_{e})$ is the contribution function for a given line and $N_{e}$ is the electron number density. By approximating the contribution function as a delta function at $T_{max}$ and assuming that the density is constant across each pixel, Equation~\ref{col_depth_one} can be written as:
\begin{equation}
4 \pi I = 0.83 G_{0} N_{e}^{2} h
\label{col_depth_two}
\end{equation}

\noindent
The {\sc eis\_density.pro} routines calculates $G_{0}$ for a given electron density which allows the value of $h$ to be derived for each pixel within a raster for which the density is known (see \citealt{youn11} for more details). The top right hand panels of Figure~\ref{eis_den_col_dep_maps} show the five column depth maps (in arcseconds) generated from the corresponding density maps from the 14 December 2007 flare (top left panels of Figure~\ref{eis_den_col_dep_maps}). Unsurprisingly, the spatial distribution of column depth closely resembles that of the density distributions, with footpoint emission exhibiting smaller column depths than the surrounding active region; less than 15$''$ in most cases, and as little as 0.01$''$ in some pixels. Similar column depths ($\sim$10~km) were found by \citealt{delz11} during a B-class flare from electron densities of 10$^{10-11}$~cm~s$^{-1}$ using Fe~XIV ratios. This was a particularly well observed event in which the ribbons were aligned in a N--S direction, parallel to the EIS slit, and were observed in 10 sequential raters.

Following from \cite{mill11}, \cite{grah13} combined the electron density and column depth measurements to estimate the column emission measure ($EM_{col}$) for a given line at a footpoint, given that $EM_{col} = \int N_{e}^{2} dh$. The bottom five panels in Figure~\ref{eis_den_col_dep_maps} show the column emission measure maps for the impulsive phase of the 14 December 2007 event. It can be seen in each of the five rasters that higher column emission measures were found at the footpoints compared to the surrounding active region, even in the Mg~VII and Si~X lines which showed no discernible density enhancements. It is also worth noting that the value of the footpoint column emission measure at 1--2~MK of $\sim$10$^{28}$~cm$^{-5}$ using this technique are comparable to those found using the regularized inversion method described by (\citealt{hann12}; right hand panel of Figure~\ref{fp_dem}).

\subsection{Continuum Emission} 
\label{ss:continuum}
Simulations by \cite{allr05} using the RADYN code of \cite{carl94,carl95,carl97}, which models the chromospheric response to both electron beam heating and backwarming from XEUV (X-ray and EUV) photons, suggested that chromospheric emission is energetically dominated by recombination (free-bound) continua, as opposed to line (bound-bound) emission. This includes the Lyman, Balmer, and Paschen continua of hydrogen, and the He~I and He~II continua. However, definitive observations of free-bound emission during solar flares have been scarce in recent years as many modern space-based instruments do not have the sensitivity, wavelength coverage, or duty cycle required to capture unambiguous continuum enhancements during flares. One exception was a serendipitous detection of increased Lyman continuum ($\sim$70\% at 910\AA) during an X-class flare via scattered light in the SUMER instrument onboard SOHO reported by \cite{lema04}. There was also an analogous case in stellar flare spectra obtained from the Extreme Ultra-Violet Explorer (EUVE) mission by \cite{chri03}. Tentative evidence for increased He~I continuum emission was also found.

\begin{figure}[!h]
\includegraphics[width=0.97\textwidth]{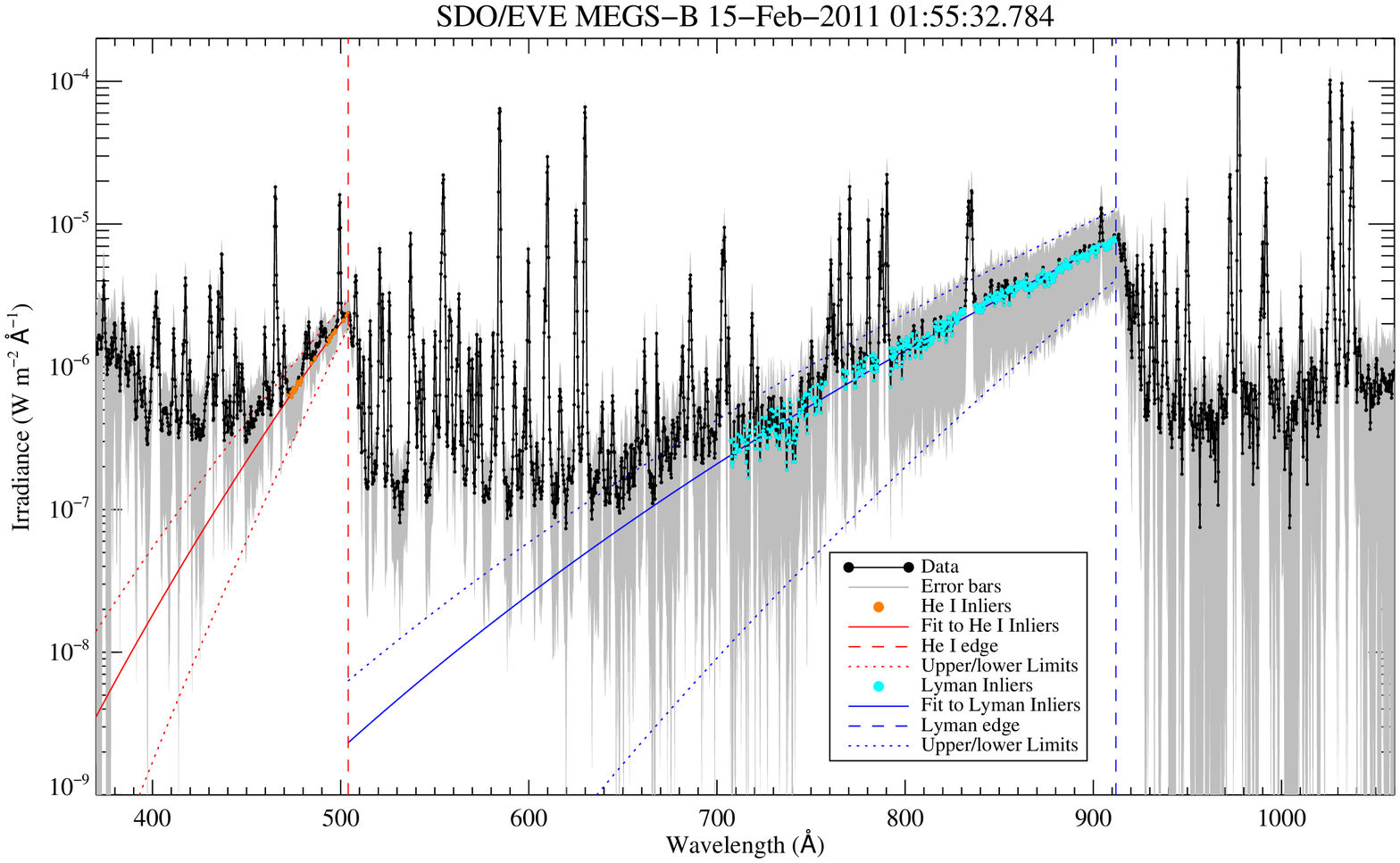}
\includegraphics[width=0.97\textwidth]{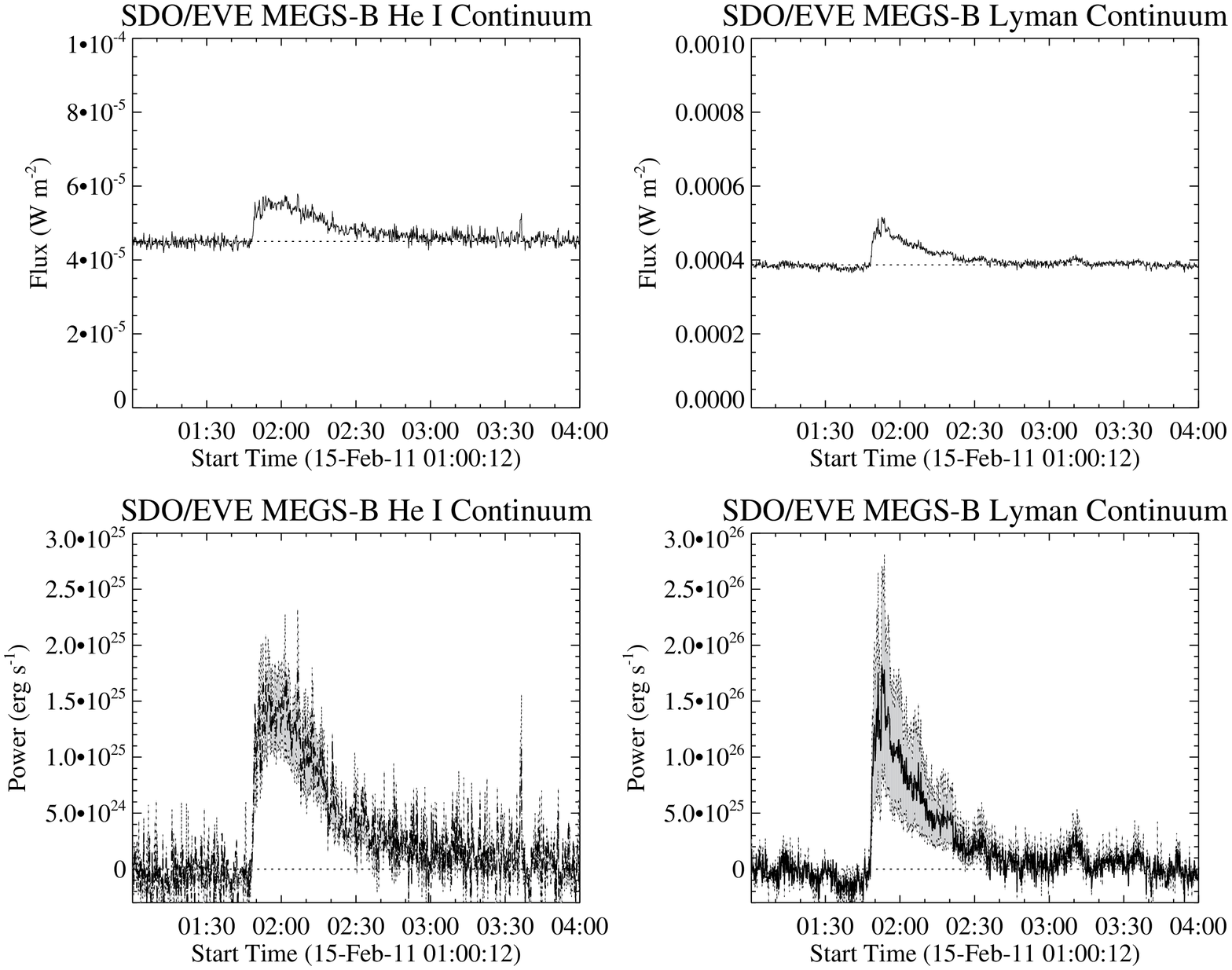}
\caption{Top panel: EVE MEGS-B spectra taken near the peak of the X2.2 flare that occurred on 15 February 2011. Overlaid in orange and cyan are the fits to the He~I and Lyman continua, respectively, using the RANSAC method \citep{fisc81}. Middle row: full-Sun lightcurves for He~I (left) and Lyman (right) continua from integrating under the fits at each 10s time step. Bottom row: Lightcurves of He~I (left) and Lyman (right) continua emission in units of energy after performing a pre-flare background subtraction. From \cite{mill14}.}
\label{eve_megsb_cont}
\end{figure}

Temporally-resolved measurements of the recombination continua of H and He have now been unambiguously observed during solar flares by \cite{mill12a,mill14}. The MEGS-B (Multiple EUV Grating Spectrograph) component of the EVE instrument, which covers the 370--1050\AA\ range (top panel of Figure~\ref{eve_megsb_cont}), includes the Lyman continuum of H~I and He~I with recombination edges at 912\AA\ and 504\AA, respectively. Using a variety of fitting techniques the time profiles of these emissions during flares were found to be impulsive (bottom four panels of Figure~\ref{eve_megsb_cont}), peaking in concert with the associated HXR emission. This suggested that the increase in continuum emission is due to recombination with free electrons that were liberated during chromospheric heating by nonthermal electrons. MEGS-A\footnote{Currently not functioning due to a power anomaly that occurred on 26 May 2014} also observed the He~II continuum with a recombination edge at 228\AA, but it is inherently weak and only observed during the largest events. It also competes with the underlying free-free continuum, as well as numerous emission lines, adding to the complexity of fitting it and deriving physically meaningful parameters. By integrating under the lightcurves displayed in the bottom row of Figure~\ref{eve_megsb_cont}, \cite{mill14} found that the Lyman and He~I continua contained $\sim$1\% and $\sim$0.1\%, respectively, of the total nonthermal electron energy as deduced from RHESSI observations during the X2.2 flare that occurred on 15 February 2011. By comparison the \lya\ and He~II 304\AA\ lines radiated away significantly more energy ($\sim$6\% and $\sim$2\%, respectively) in contrast to the predictions of \cite{allr05}.

Although energetically very weak, these continua serve as powerful diagnostic tools for determining the depth in the solar atmosphere at which they are formed. Following from \cite{mach78}, \cite{ding97} stated that the temperature of the continuum (relative to a blackbody) can be ascertained using:

\begin{equation}
I_{\lambda} \approx \frac{1}{b_1} \frac{2hc^2}{\lambda^5} exp\left( \frac{-hc}{\lambda kT}\right)
\end{equation}

\noindent
where the departure coefficient, $b_1$, is given by:
\begin{equation}
\frac{2hc^2}{I_{\lambda_1}\lambda_1^5}exp\left(\frac{-hc}{\lambda_1 kT}\right) = \frac{2hc^2}{I_{\lambda_2}\lambda_2^5}exp\left(\frac{-hc}{\lambda_2 kT}\right)
\end{equation}

\noindent
Rearranging gives:

\begin{equation}
T = \frac{hc}{k}\left(\frac{1}{\lambda_1}-\frac{1}{\lambda_2}\right)\left[ln\left(\frac{I_{\lambda_2}\lambda_2^5}{I_{\lambda_1}\lambda_1^5}\right)\right]^{-1}
\end{equation}

EVE measurements of flare-related increases in Lyman continuum emission carried out in conjunction with observations in the Balmer and Paschen continua as recently reported by \cite{hein14} and \cite{kerr14}, respectively, would greatly advance our knowledge of the distribution of energy throughout the lower solar atmosphere, as well as the dominant emission mechanism, particularly when compared to radiative hydrodynamic simulations such as those from \cite{allr05}.

\section{\lya\ Flare Observations}
\label{lya_flares}

Although the \lya\ line of H~I at 1216\AA\ straddles the boundary between the EUV and UV portions of the spectrum, its relevance to chromospheric flare heating and the current availability of photometric \lya\ data from EVE MEGS-P warrants a brief summary of past observations and the potential for future diagnostics. True spectroscopic observations of \lya\ have been few and far between lately largely due to the technical challenges inherent in measuring the brightest line in the solar spectrum. Among the first spectrally-resolved \lya\ flare observations were carried out using the S082B spectrograph onboard Skylab, although saturation was often an issue \citep{lite79,canf80,doyl92}. Prior to this, \lya\ photometers had been flown on sounding rockets and early satellites but no flare-related enhancements were detected \citep{chub57,krep62}. The Solar Stellar Irradiance Comparison Experiment (SOLSTICE) instrument onboard the Solar Radiation and Climate Experiment (SORCE) satellite typically takes spectral irradiance measurements across the EUV range once per orbit. However, during the Halloween flares of October 2003 it was fortuitously scanning through the \lya\ line at 1~minute cadence. \cite{wood04} reported only a 20\% increase in the core of the line during the 28 October 2003 X28 flare, but the wings of the line were found to increase by a factor of two. This is significantly higher than the 6\% increase reported by \cite{brek96} during the decay phase of an X3 flare on 27 February 1992 using the same instrument. \cite{john11} were able to detect enhancements in \lya\ luminosity from scattered light using off-limb observations from SOHO/Ultraviolet Coronagraph Spectrometer (UVCS), while \cite{rubi09} estimated that \lya\ emission accounts for $<$10\% of a flare's radiated energy budget using broadband measurements from the Transition Region And Coronal Explorer (TRACE; \citealt{hand99}). Flare-related enhancements in \lya\ in EVE data have also been reported by \cite{mill12a} for the 15 February 2011 X2.2 flare, and a follow-up study also revealed \lya\ emission to contribute 6-8\% of the total measured radiated losses from the chromosphere \citep{mill14}. In contrast, \cite{kret13} found only a 0.6\% increase in \lya\ emission during an M2 flare using the LYRA radiometer onboard PROBA2. More strikingly though was that its temporal behaviour appeared to mimic that of the GOES SXR emission, rather than of the impulsive HXR emission as one might expect for intrinsic chromospheric emission. They found that by taking the time derivative of the \lya\ profile, the resulting peaks closely resembled those of the derivative of the SXR lightcurve. The authors cited this as evidence for pre-flare heating prior to the onset of accelerated particles. The possibility of gradual heating by thermal conduction cannot be ruled out but it is more likely that instrumental factors were contributing to this anomalous behaviour (see below). \cite{kret13} also provides a brief overview of many recent \lya\ flare observations. 

\begin{figure}[!t]
\includegraphics[width=\textwidth]{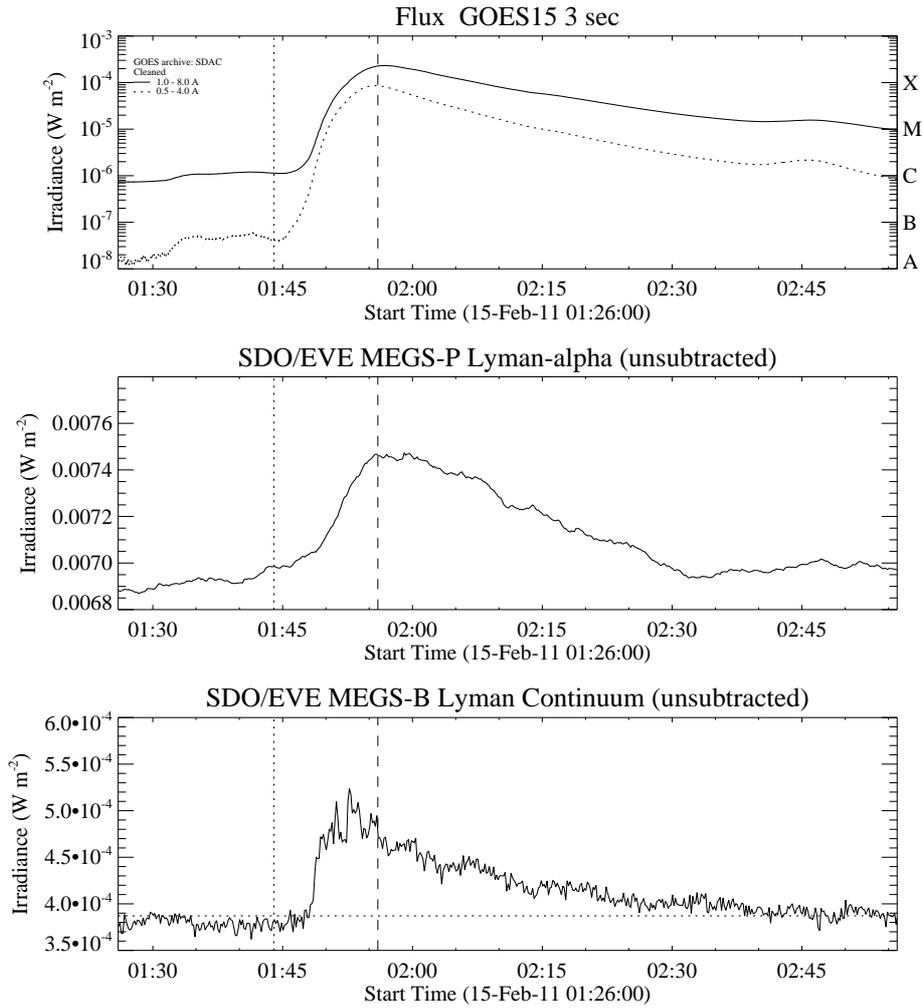}
\caption{Top panel: GOES SXR lightcurves during the X2.2 flare that occurred on 15 February 2011 in 1--8\AA\ (solid curve) and 0.5--4\AA\ (dotted curve). Middle panel: \lya\ lightcurve taken using SDO/EVE MEGS-P for the same event. Bottom panel: LyC lightcurve from SDO/EVE MEGS-B data formed using the techniques described in Section~\ref{ss:continuum}. The vertical dotted and dashed lines in each panel denote the start and peak times of the flare, respectively, as stated in the GOES event list.}
\label{eve_lyman_ltc}
\end{figure}

SDO/EVE now provides routine, broadband (100\AA) observations of flares in \lya\ through its MEGS-P diode. Despite the lack of spectral and spatial information, these data are useful in terms of quantifying the energy budget of radiative losses in the chromosphere. However, similar to the temporal behaviour found by \cite{kret13}, the lightcurves of \lya\ from EVE appear to show a more gradual, ``GOES-like'' profile. Figure~\ref{eve_lyman_ltc} shows the SXR (top), \lya\ (middle), and LyC (bottom) time profiles for the 15 February 2011 X-class flare. While the LyC profile appears impulsive as one might expect for a chromospheric plasma heated via Coulomb collisions, \lya, in contrast, displays a rise time of 10-20 minutes, akin to that seen in SXR. The reason for this disparity is unclear at present. The spectrally-resolved \lya\ profiles shown by \cite{wood04} for the 28 October 2003 flare using SORCE/SOLSTICE peaked in sync with the derivative of the SXR, confirming an impulsively heated atmosphere. This suggests that the broadband nature of the EVE (and LYRA) measurements are in some way ``smoothing out'' this intrinsic, bursty nature. Perhaps other, higher temperature lines or continua are contributing to the passband, or a data processing algorithm in the EVE pipeline is responsible. Once this issue is resolved, simultaneous \lya\ and LyC observations (along with \lyb, \lyc, etc) will be a valuable tool for investigating heating of the lower solar atmosphere, particularly in conjunction with IRIS.

\section{Concluding Remarks}
\label{s:conc}
EUV spectroscopy has advanced considerably over the past 50 years or so. However, its application to flares, particularly at footpoints and ribbons during the impulsive phase, has been somewhat lacking given the diagnostic potential available. The studies referenced in this paper highlight the capabilities (and limitations) of our current spectrometers, and the scientific accomplishments that have been achieved. (For other reviews on chromospheric flares, see \citealt{huds07,flet11,flet12}, and \citealt{flet13}.) However, many of these works have been carried out in isolation, or perhaps in conjunction with imaging data for context information. I believe that the greatest advances will be made by combining datasets from both EIS and EVE together with those from other instruments such as SDO/AIA, Hinode/SOT, RHESSI, IRIS, and ground-based facilities like those at the Dunn Solar Telescope (DST). On 29 March 2014, an X1.0 flare was observed by all of these instruments making it one of the best observed flares ever, and which will no doubt yield some interesting results as well as validating the advantage of coordinated observations. October 2014 also saw a hugely successful coordinated observing campaign facilitated by the Max Millennium Program for Solar Flare Research\footnote{\url{http://solar.physics.montana.edu/max_millennium/}} in support of Service Mode Operations at the DST with coordinated observations from RHESSI, SDO/EVE MEGS-B, IRIS, and all three Hinode instruments. These Service Mode Operations are how flare observations are anticipated to be run at the Daniel K. Inoue Solar Telescope (DKIST; formally the Advanced Technologies Solar Telescope - ATST). Multi-wavelength campaigns such as these should be strongly encouraged.

As we enter the declining phase of Solar Cycle 24, the opportunities for coordinated campaigns may be somewhat limited. However, both EIS and EVE have several years of data waiting to be explored and exploited (see \citealt{wata12} for a complete list of flares observed by EIS to date and a similar list\footnote{\url{http://lasp.colorado.edu/eve/data_access/evewebdata/interactive/flare_campaign_observations.html}} has been compiled for all flares observed during EVE MEGS-B campaigns). This will help us move away from looking at individual events and begin investigating the collective properties of a statistically significant number of events. EVE data may also permit a comparison between flares in solar and stellar atmospheres by looking at analogous events observed by EUVE, for example, which covered a similar spectral range. But perhaps the most meaningful collaborations will be those that attempt to model the observed EUV features and their derived parameters through numerical simulations. Many current flare heating models are now capable of recreating atmospheric conditions such as temperature and density, as well as optically thin emission lines themselves, in response to an injection of energy \citep[e.g.][]{brad03,klim08,liu09,kasp09}. This injection can, and should, be based on nonthermal electron parameters derived from RHESSI (or Fermi) data with the model output being directly compared with solar observations. Any serious discrepancies between predictions and reality may mean considering other forms of energy transport, such as thermal conduction \citep{bran14} or Alfv\'{e}n waves \citep{russ13}. Modelling optically thick emission adds another layer of complexity in the form of solving the equations of radiative transfer \citep{allr05}. Ultimately, a comparison between the outputs of chromospheric heating models and the corresponding EUV observations - based on a common energy input - is needed to gain a greater insight into the driving mechanism behind enhanced flare emission, at all wavelengths, in the lower solar atmosphere.

\section{Future Prospects}
\label{s:future}
The continued operation of EIS and EVE into the decay of Solar Cycle 24 promises that more interesting discoveries lie ahead. In the longer term, EUV spectroscopy remains a high priority for future missions. The current plan for Solar-C (the successor to Hinode) features an imaging spectrometer similar to EIS called the EUV Spectroscopic Telescope (EUVST). By performing spectral analysis of emission lines, EUVST will measure the velocity, temperature, and density of solar plasmas. It will have more than 5 times the spatial resolution and 10 times the throughput of the currently operational spectrometers, and will be able to observe over a wide wavelength range (170\AA\ to 1280\AA), seamlessly covering the transition region from the chromosphere to the coronal and extend to flare temperatures. A high throughput performance will enable spectroscopic observations with short exposure times of around 1~s, so that it can reliably capture the rapid temporal fluctuations of magnetic structures and waves resulting from magnetic reconnection, jets, and photospheric motions.

Solar Orbiter, due to launch in 2018, will comprise the Spectral Imaging of the Coronal Environment (SPICE) instrument, an EUV imaging spectrograph designed to observe both on the solar disk and out in the corona to remotely characterise plasma properties at and near the Sun. Specific scientific topics to be addressed by SPICE include studies of solar wind origin by matching {\it in situ} composition signatures in solar wind streams to surface feature composition; studies of the physical processes that inject material from closed structures into solar wind streams; and studies of SEP source regions by imaging remotely the suprathermal ions thought to be seed populations of SEPs. The SPICE observing strategy is to produce 2D spectroheliograms of selected line profiles and line intensities. Like EUVST, the selected lines represent the full range of temperatures and heights in the solar atmosphere, from the chromosphere to the flaring corona. Solar Orbiter will also carry an EUV Imager (EUI) that will comprise a \lya\ channel. Resolving the discrepancies with current \lya\ observations is therefore paramount to insure future measurements are interpreted correctly, particularly during explosive events.

With Solar Cycle 25 set to peak around 2025, the prospect of coordinated observations between Solar-C and Solar Orbiter, along with Solar Probe Plus and the DKIST, will mark an auspicious time to further our understanding of the most dynamic, complex, and perplexing region of the Sun's atmosphere, and that of other stars. Given the importance of the chromosphere and lower transition region to flare energetics, the acceleration of the solar wind, and the overarching field of space weather, diagnosing this crucial layer should be a primary focus of future missions.

\begin{acks}
The author is grateful for financial support from NASA for LWS/TR\&T grant NNX11AQ53G and LWS/SDO Data Analysis grant NNX14AE07G. He also thanks Drs. Brian Dennis, Mihalis Mathioudakis, Peter Young, Hugh Hudson, Helen Mason, and David Graham for their useful comments and feedback on the manuscript, as well as the organisers of the meeting on Solar and Stellar Flares: Observations, Simulations, and Synergies in Prague, Czech Republic in June 2014 for the invitation to present this review.
\end{acks}

\bibliographystyle{spr-mp-sola}
\bibliography{ms}  

\end{article} 
\end{document}